\documentclass[12pt]{article}
\title{Coleman's theorem on physical assumptions \\
for no Goldstone bosons in two dimensions}
\author{Chigak Itoi \\ 
Department of Physics,  Nihon University, \\
Kanda-Surugadai, Chiyoda, Tokyo 101-8308, Japan} 
\begin{document}
\maketitle
\begin{abstract}
Thirty years ago, Coleman proved
that no continuous symmetry is broken spontaneously
in a two-dimensional relativistic quantum field theory.
In his argument, however, it is difficult 
to understand the physical meaning of the assumption 
of no infrared divergence.
I derive the same result directly from the cluster property
of a local field regarded as a physically 
acceptable assumption.
\end{abstract}
%\pacs{11.10.Kk, 11.30.Qc}
\paragraph{\it  Introduction.}
Coleman's theorem claims that the 
vacuum expectation value of a charged field 
must vanish in a two-dimensional
relativistic quantum field theory with 
a continuous symmetry \cite{C}.
This is a remarkable nature of relativistic field theories,
while in some non-relativistic systems in one space dimension
a continuous symmetry is broken spontaneously
at zero temperature. 
This theorem is quite useful so that his renowned 
paper has been 
cited many times for thirty years. 
In a few articles \cite{N,FI},
however, one finds criticism to Coleman's theorem 
based on some counter examples.
Always, the assumption for no Goldstone bosons
comes into question, in addition to an incorrect representation
of Coleman's lemma.
Here, I discuss a physical 
problem in the assumption of the theorem. 
Coleman assumed that a field theory with an infrared 
divergence 
in Wightman functions is ill-defined and therefore 
no massless scalar fields exist in two dimensions.
Without this assumption
one cannot show the theorem.
The title of his paper 
``There are no Goldstone bosons in two dimensions" 
is not the result, but an assumption in the theorem.
Generally in physics, however, infrared divergence is 
important phenomenon.
Therefore, the existence of infrared divergence 
should not be a criterion to exclude ill-defined theories.
For example, the ferromagnetic quantum Heisenberg model
in one space dimension has a divergent magnetic susceptibility 
at zero temperature and the SU(2) symmetry is broken spontaneously. 
In this case, the infrared divergence helps the ferromagnetic order 
to compete with the quantum fluctuation. Generally
for a locally interacting lattice system
with spontaneously broken continuous symmetry in one space dimension,
Momoi proved that the corresponding susceptibility
should diverge at zero temperature \cite{M}.
The infrared divergence plays a physically important role in this case.
On the other hand, 
an infrared divergence is not so serious in some other cases.
For example, one can use a massless scalar field 
in two dimensions to bosonize a fermion field or 
some others with different scaling dimensions. 
In this case, the infrared divergence can be easily 
removed from the physically important correlation functions.
Therefore, the massless scalar field is not ill-defined even in
two dimensions.  
The mathematical 
assumption to exclude massless scalar fields because
of no infrared divergence in Coleman's argument
is not understandable physically. 

In this paper, 
I show impossibility of non-zero 
vacuum expectation value only for a local scalar 
field with a cluster property. 
This property is necessary for a 
local field above a healthy vacuum and 
sufficient for the theorem in any subtle case.
Although the cluster property 
of a local field is proved in  
a primary representation of a $C^*$-algebra \cite{H}, 
it is a physical requirement for local fields
from the viewpoint of physicists.  
The theorem proved on this assumption 
is indeed sufficiently useful for physical systems.
It is neither surprising nor problematic
to obtain a non-zero vacuum 
expectation value of a scalar field with no cluster property 
in two dimensions. All examples \cite{N,FI} with a non-zero 
vacuum expectation 
value of scalar fields  
are not true counter examples to 
the theorem from this viewpoint. 
In the following, I present my results along Coleman's style. 
\paragraph{\it Assumptions.} 
A current $j_\mu(x)$ satisfies 
$\partial^{\mu} j_\mu (x) =0$
for a continuous symmetry. A 
real scalar field $\phi_i(y) \ (i=1, 2, \cdots)$ is transformed
by the current through an equal-time commutation relation
\begin{equation}
[j_0(x), \phi_i(y)]|_{x_0=y_0} = \sum_k C_{ik}\phi_k(x) \delta(x_1-y_1).
\label{commutation}
\end{equation}
The charge 
$
Q \equiv \int dx_1 j_0(x_0, x_1)
$
is a generator of the global symmetry. 
The present argument can be apply to a charge in a set of several charges
$Q_i \ (i=1, 2, \cdots)$. %and for fields $\phi_j(x)$ satisfying
%a homogeneous relation $[Q_i, \phi_j(x)]= C_{ij} ^k \phi_k(x)$.
I assume also the relativistic covariance of the theory.
I will show that the vacuum expectation value 
of the charged local field 
must vanish if the correlation functions of this field 
satisfy the cluster property.
The cluster property of the two-point function requires
\begin{equation}
\langle 0 | \delta \phi_i(x) \delta \phi_i(y) | 0 \rangle 
\rightarrow 0,
\label{cluster}
\end{equation}
as a space-like separation $(x-y)^2 \rightarrow -\infty$, 
where the deviation is defined by
$\delta \phi_i(x) \equiv \phi_i(x)  
-\langle 0 | \phi_i(x) | 0 \rangle$.
For example in a free field representation in 
unperturbed or perturbed conformal field theories, 
a compact boson $\varphi(x)$ itself does not have the cluster property.
On the other hand, a physical field operator with a positive conformal dimension, 
say $e^{2 \pi i \varphi/R}$
with a compact boson radius $R$ or
a current $\partial_\mu \varphi(x)$,
has the cluster property. The infrared divergence 
can be easily removed from correlation functions
of the physical operators by some suitable rescaling. 
Hereafter, we assume that the physically meaningless 
divergence is removed.
The Fourier transformed two point functions are represented in
\begin{eqnarray}
%&&F( k ) = 
&& \int \!d^2 x e^{i k x}
\langle 0 |  \delta \phi_i(x) \delta \phi_i(0)  | 0 \rangle=
\rho_{ii}(k^2) \theta(k_0), \label{scalar} \\
%&&F_\mu( k ) = 
&& \int \! d^2 x e^{i k x}
\langle 0 | \delta j_\mu (x)\delta \phi_i(0) | 0 \rangle
=(\sigma_i k_\mu  \delta(k^2) + 
\epsilon_{\mu\nu}  k^\nu \epsilon (k^1) \rho_i(k^2))\theta(k_0) , \label{vector} \\
%&&F_{\mu \nu}( k ) = \! 
&&\int \! d^2 x e^{i k x}
\langle 0 | \delta j_\mu(x) \delta j_\nu(0) | 0 \rangle=
(k_\mu k_\nu-\eta_{\mu \nu}k^2)\rho(k^2) \theta(k_0). 
\label{tensor} 
\end{eqnarray}
by using the relativistic covariance, the positivity of energy,  
the permutation symmetry and the current conservation.
Note that the sign function $\epsilon(k^1)$ in Eq.(\ref{vector}) Coleman overlooked 
is necessary for the covariance under the parity transformation.  
I assume the spectral condition $\rho_{ii}(k^2)=\rho_i(k^2)=\rho(k^2)=0 $ for $k^2 < 0$,
and the positive semi-definiteness 
\begin{equation}
\rho_{ii}(k^2) \geq 0, \ \ \ \ \ \rho(k^2) \geq 0,
\label{positive}
\end{equation}
and also I assume their integrability over $0 < k^2 < \infty$.
The following expression of a commutator 
is convenient for later discussions  
\begin{eqnarray}
%\int d^2 x e^{i k x} \langle 0 | \{ \delta \phi_i(x), 
%\delta \phi_i(0)  \}| 0 \rangle &=&\rho_{ii}(k^2), \label{scalar2} \\
 \int d^2 x e^{i k x} \langle 0 | [\delta j_\mu (x), \delta \phi_i(0)  ] | 0 \rangle
&=& %\frac{1}{2}(
 (\sigma_{i1}%+ \sigma_i^*) 
k_\mu   + 
%\frac{1}{2}(
i \sigma_{i2}%-\sigma_i^*) 
k_\mu  \epsilon(k_0)  
 + i \sigma_i' \epsilon_{\mu\nu}  k^\nu \epsilon (k^1)) \delta(k^2) \nonumber \\
&+&  \epsilon_{\mu\nu}  k^\nu \epsilon (k^1) \epsilon(k_0) 
\rho_{i1}(k^2), \label{vector2} 
%\int \! d^2 x e^{i k x}\langle 0 |\{ \delta j_\mu(x),\  
%\delta j_\nu(0) \}| 0 \rangle &=&(k_\mu k_\nu-\eta_{\mu \nu}k^2)\rho(k^2), \label{tensor2} 
\end{eqnarray}
where $\sigma_i = \sigma_{i1} + i \sigma_{i2}$ and  $\rho_{i}(k^2)=
\rho_{i1}(k^2)+i \rho_{i2}(k^2)$.
The function $\rho_{i2}(k^2)$ has been replaced by $\sigma' \delta(k^2)$ 
because of the commutation relation Eq.(\ref{commutation}).
Then the vacuum expectation value is
written in
\begin{equation}
\sum_k C_{ik}\langle 0| \phi_k(0)|0 \rangle = \langle 0|\int d x_1 
[j_0(x_0, x_1),\phi_i(0)]|0 \rangle = 
\frac{i}{2 \pi} (\sigma_{i2}+ \sigma_i').
\label{VEV}
\end{equation}
Note this pure imaginary valued equation different from 
the complex $\sigma_i$ in the Wightman function (\ref{vector}).
Because of this difference, one cannot employ Coleman's  
inequality for the Wightman functions to prove the theorem.   
\paragraph{\bf Lemma.}
{\it Let $H(\kappa^2)$ be a positive semi-definite function 
of a single variable $\kappa^2 \geq 0$ 
with a bound $H(\kappa^2) \leq C G(\kappa^2)$ 
by some constant $C > 0$ and
a function 
\begin{equation}
G(\kappa^2) \equiv \int_{-\infty} ^\infty dp
\frac{\cos p}{\sqrt{p^2+\kappa^2}}.
\label{G}
\end{equation} 
For the distribution function $\rho_{ii}(k^2)$ 
in Eq.(\ref{scalar}) of a 
field $\phi(x)$ 
with the cluster property (\ref{cluster}),
the following limit 
vanishes}
\begin{equation}
\lim_{\lambda \rightarrow \infty} \int_0 ^\infty \frac{d \kappa^2}{\lambda}
\rho_{ii} \left(\frac{\kappa^2}{\lambda}\right) H(\kappa^2) =0.
\label{lemma}
\end{equation}
\paragraph{\it Proof of Lemma.}
A spectral representation of the Fourier transformed two-point function 
is obtained by changing the integration variable from $k_0$ to $m^2= k_0^2-k_1^2$. 
In this representation, the cluster property (\ref{cluster}) gives
\begin{equation}
\lim_{|x_1| \rightarrow \infty}
\int_0 ^\infty dm^2 \rho_{ii}(m^2) \int_{-\infty} ^\infty
dk_1 \frac{ \cos k_1 x_1}{\sqrt{k_1^2+m^2}} =0.
\label{cluster2}
\end{equation}
By rescaling the integration variables, this condition is rewritten into
\begin{equation}
\lim_{|x_1| \rightarrow \infty}
\int_0 ^\infty \frac{d\kappa^2}{|x_1|^2} 
\rho_{ii}\left(\frac{\kappa^2}{|x_1|^2}\right) G(\kappa^2)=0,
\label{cluster3}
\end{equation}   
By the bound of $G(\kappa^2)$ and the positive semi-definiteness 
of each function,
the right hand side of Eq.(\ref{lemma}) is bounded by
\begin{equation}
\int_0 ^\infty \frac{d \kappa^2}{\lambda} \rho_{ii}\left( \frac{\kappa^2}{\lambda} \right)
H(\kappa^2) \leq C \int_0 ^\infty \frac{d\kappa^2}{\lambda} 
\rho_{ii}\left(\frac{\kappa^2}{\lambda}\right) G(\kappa^2).
\end{equation}
The right hand side
vanishes in the limit with the correspondence $\lambda=|x_1|^2$ 
in Eq.(\ref{cluster3}).    
Thus the lemma has been proved.
\paragraph{\bf Theorem.}
{\it The vacuum expectation value of a charged scalar field with 
the cluster property vanishes in a two-dimensional
relativistic quantum field theory with a conserved charge 
with respect to the continuous symmetry. }
\paragraph{\it Proof of Theorem.}
I evaluate terms in an uncertainty relation in the vacuum,
instead of Coleman's correlation inequality
\begin{equation}
\langle 0|  \delta A(\lambda)^2|0 \rangle 
\langle 0| \delta B(\lambda)^2|0 \rangle
\geq \left|\frac{1}{2}\langle 0|  [A(\lambda), B(\lambda)] |0 \rangle \right|^2,
%[\int d^2 k  |h_\lambda(k)|^2][ \int d^2 k F_{00}(k) |h_\lambda(k)|^2]
%\geq \left|\int d^2k F_0(k)|h_\lambda(k)|^2 \right|^2, 
\label{inequ}
\end{equation}
for operators 
$$A(\lambda) = \int d^2x h_\lambda(x) \phi_i(x), 
\ \ \  \ B(\lambda) = \int d^2x h_\lambda(x) j_0(x).
$$
Here, I choose a test function 
$$
h_\lambda(x) = \int \frac{d^2k}{(2 \pi)^2} e^{ikx} \tilde{h}_\lambda(k), \ \ \  
\tilde{h}_\lambda(k) \equiv
f(\lambda k^2) g(k_1^2),
$$ 
by bounded positive semi-definite functions $f(\kappa^2)$ and $g(\kappa^2)$
with a compact support. The support of $f(\kappa^2)$ 
includes the origin and that of $g(\kappa^2)$
does not. For example, 
one of the simplest choice is
$f(\kappa^2)= \theta(a^2-\kappa^2)$
and $g(k_1^2)=\theta(k_1^2-b^2)\theta(c^2-k_1^2)$ with 
positive numbers  $a < b < c$.
This test function
differs from the original one, but has the same
important property. The support of $\tilde h_\lambda(k)$  
lies on the light cone in the large $\lambda$ limit. 
Now, I evaluate the first term in the inequality (\ref{inequ})
\begin{eqnarray}
%\int d^2 k F(k) |h_\lambda(k)|^2
\langle 0|  \delta A(\lambda)^2|0 \rangle 
 &=&\int_0 ^\infty \!\!dm^2 \rho_{ii}(m^2) | f(\lambda m^2)|^2
\int_{-\infty} ^\infty \!\!\!\!dk_1 
\frac{ |g( k_1^2)|^2}{ 2 \sqrt{k_1^2+m^2}} 
\nonumber \\ 
&\leq&\int_0 ^{\infty} \frac{d\kappa^2}{\lambda} 
\rho_{ii}\left(\frac{\kappa^2}{\lambda}\right) |f(\kappa^2)|^2 
\int_{-\infty} ^\infty \!\!\!\!dk_1 
\frac{ |g( k_1^2)|^2}{2 |k_1|}.
\nonumber
\end{eqnarray}
Since 
the support of the function $g(k_1^2)$ does not include
the origin, the factor $ \int dk_1 
|g( k_1^2)|^2/|k_1|$ is finite.
The bounded positive semi-definite function $|f(\kappa^2)|^2$ 
with a compact support is bounded by $C G(\kappa^2)$ 
with some positive constant $C$, 
then the first term in the inequality (\ref{inequ}) 
vanishes in the limit by the lemma 
\begin{equation}
\lim_{\lambda \rightarrow \infty} \langle 0| \delta A(\lambda)^2 
|0 \rangle 
%\int d^2 k F(k) |h_\lambda(k)|^2
=0. 
\label{limitscalar}
\end{equation}
Next, I evaluate the second term in the left hand side of (\ref{inequ})  
\begin{equation}
%\int d^2 k F_{00}(k) |h_\lambda(k)|^2 
\langle 0| \delta B(\lambda)^2 |0 \rangle 
 = \int_0 ^{\infty} \!\!\!\!dm^2 \rho(m^2) |f(\lambda m^2)|^2
\int_{-\infty} ^\infty dk_1 
\frac{k_1^2 |g(k_1^2)|^2}{2 \sqrt{k_1^2+m^2}} 
\label{limitensor}
\end{equation}
which is also bounded from above because of the boundedness 
of the integrand and the integrability
of $\rho(m^2)$. Finally, I show that
the right hand side of (\ref{inequ}) is proportional
to the square of the vacuum expectation value with a non-zero coefficient. 
The $\delta$-functional singularity in Eq.(\ref{vector2}) gives
\begin{equation}
%\int d^2k F_0(k)|h_\lambda(k)|^2 
\langle 0|  [A(\lambda), B(\lambda)] |0 \rangle 
 = i (\sigma_{i2}+ \sigma_i') f(0) \int_{-\infty} ^\infty dk_1 |g(k_1^2)|^2 .
\label{limitvecor}
\end{equation}
Therefore, %Coleman's correlation inequality 
the uncertainty relation (\ref{inequ}) 
in the large $\lambda$ limit prohibits the non-zero vacuum 
expectation value (\ref{VEV}) of the charged local field with the cluster property. 
The theorem has been demonstrated in more physical situation.

\paragraph{\it Note.}
In this paper, the theorem is shown by a test function with 
a compact support including the light cone in a momentum space
as in Coleman's proof. This corresponds to setting a test state
extended over the whole space-time.
To prove the theorem with a test state
on a space-like surface as in
the famous Mermin-Wagner theorem \cite{MW}, 
Bogoliubov's inequality is no use at zero temperature. 
One can use more complicated Shastry's inequality, which was
employed to prove no antiferromagnetic order
in the antiferromagnetic quantum Heisenberg chain \cite{S}. 
This inequality %(\ref{inequ}) 
works with a test function $\tilde h_\lambda (k) = g(\lambda k_1^2)$, 
where the $k_0$-independence implies no support compactness
and a test state on a space-like surface. 
In this case, one needs the lemma 
under a relaxed condition on $H(\kappa^2)$ 
when the bound by $G(\kappa^2)$ is available only in some 
neighborhood of the origin.
This version of lemma 
allows one to prove the theorem with the 
test state on a space-like surface.
\paragraph{\it Acknowledgements}{\small The author
would like to thank T. Nihei and M. Yamanaka for reading the manuscript 
and helpful comments. He is grateful to M. Homma for helpful 
discussions.
He would like to thank M. Faber and A. N. Ivanov for their point out 
errors in the manuscript.}

\end{document}